\documentclass[aps,reprint,showpacs,pra,superscriptaddress]{revtex4-1} 
\usepackage{graphicx}
\usepackage{amssymb}
\usepackage{amsmath}
\usepackage{bm}
\usepackage{hyperref}

\newcommand{\w}{\omega}    
\newcommand{\I}{i}

\begin{document}

\title{Doppler-free coherent-control spectroscopy with a colliding pair of shaped pulses}

\author{Minhyuk~Kim}
\author{Kyungtae~Kim} 
\affiliation{Department of Physics, KAIST, Daejeon 305-701, Korea}

\author{Dewen~Cao}
\affiliation{Institute of Intelligent Machines, Chinese Academy of Sciences, Hefei, Anhui 230031, China}
\affiliation{Department of Automation, University of Science and Technology of China, Hefei, Anhui 230026, China}

\author{Fang~Gao}
\affiliation{Institute of Intelligent Machines, Chinese Academy of Sciences, Hefei, Anhui 230031, China}

\author{Feng~Shuang}
\affiliation{Institute of Intelligent Machines, Chinese Academy of Sciences, Hefei, Anhui 230031, China}
\affiliation{Department of Automation, University of Science and Technology of China, Hefei, Anhui 230026, China}
\affiliation{Department of Mechanical Engineering, Anhui Polytechnic University, Wuhu Anhui 241000, China}

\author{Jaewook~Ahn}
\affiliation{Department of Physics, KAIST, Daejeon 305-701, Korea}
\email{jwahn@kaist.ac.kr}

\date{\today}

\begin{abstract}
We demonstrate the use of the ultrafast spatial coherent-control method to resolve the fine-structure two-photon transitions of atomic rubidium. Counter-propagating ultrafast optical pulses with spectral phase and amplitude programmed with our optimized solutions successfully induced the two-photon transitions through $5S_{1/2}$-$5P_{1/2}$-$5D$ and $5S_{1/2}$-$5P_{3/2}$-$5D$ pathways, both simultaneously and at distinct spatial locations. Three different pulse-shaping solutions are introduced that combine amplitude shaping, which avoids direct intermediate resonances, and phase programming, which enables the remaining spectral components to be coherently interfered through the targeted transition pathways. Experiments were performed with a room-temperature vapor cell, and the results agree well with theoretical analysis. 
\end{abstract}
%\keywords{}

\maketitle

\section{Introduction}

Light structured in the spectro-temporal domain is used in coherent control to enhance or suppress nonlinear material responses through engineered passages of optical transitions~\cite{Bergmann,Shapiro,Tanner}. It has been demonstrated that laser pulses with programmed phase and amplitude can boost, for example, the two-photon transitions of atoms in two, three, and four energy-level structures~\cite{Dudovich, SLee, MCStowe,LeePRA2013}. Various functional light-matter interactions have also been designed with shaped lights; examples include dark pulses~\cite{Dark_pulses}, laser-catalytic chemical reactions~\cite{Teaching_laser}, quantum gates~\cite{Quantum_gates}, and high-harmonic generations~\cite{HHG}, to list a few. In particular, state-to-state controllability in coherent control allows for the selective excitation of otherwise unresolvable energy states of a complex quantum system~\cite{NDudovich}, promising the use of coherent control methods in precision and/or functional spectroscopy. 

The use of shaped-pulses in laser spectroscopy became more interesting following the recent demonstration of Doppler-free coherent-control spectroscopy~\cite{BarmesNPhoton2013}. Termed as ultrafast spatial coherent-control (USCC), this method uses a pair of colliding laser pulses, of which the spectrum is phase modulated in such a way that counter-propagating photons that satisfy each two-photon excitation meet at a specific location along the beam propagation path. With this method, laser pulses were successfully programmed to provide a spatially-mapped spectroscopic assessment of each two-photon transition of rubidium and cesium all at once~\cite{BarmesNPhoton2013}. In particular, in conjunction with the optical frequency comb~\cite{JonesScience2000, HolzwarthPRL2000}, USCC holds a promise for ultra-high precision spectroscopy of atomic species for which the laser cooling method is unavailable~\cite{BarmesPRL2013}.

\begin{figure}
\includegraphics[width=1.0\columnwidth]{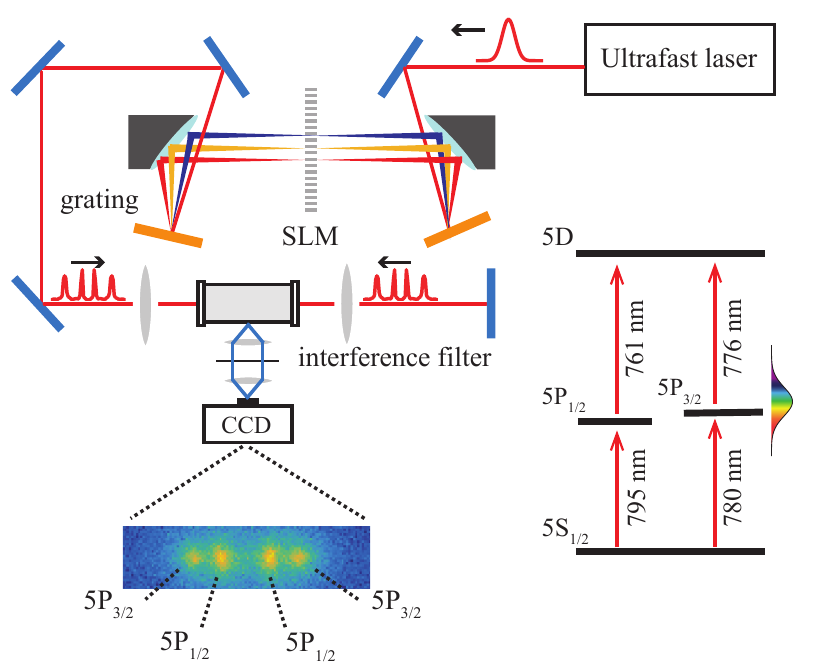}
\caption{Ultrafast spatial coherent-control scheme: Laser pulses are phase-modulated to spatially resolve the Doppler-free excitation of $5S_{1/2}$-$5P_{1/2}$-$5D$ and $5S_{3/2}$-$5P_{1/2}$-$5D$ of atomic rubidium.}
\label{fig1}
\end{figure}

Our previous work extended this method~\cite{BarmesNPhoton2013} to a two-photon transition system having an intermediate resonance using a special spectral-phase modulation~\cite{WJLeePRA2015}.  
In this paper, consiering the fact that multiple transition pathways may exist in quantum systems~\cite{Gao2014,Gao2015,Gao2016}, we extend this approach~\cite{WJLeePRA2015} even further to a system of multiple transition pathways, where the crowding  resonances require special spectral-phase modulations. 
We propose three such modulations, particularly designed to resolve the fine-structure energy levels of alkali atoms and experimentally demonstrate their performance. The experimental concept is shown in Fig.~\ref{fig1}, where the two fine-structure two-photon transition pathways, $5S_{1/2}$-$5P_{1/2}$-$5D$ and $5S_{3/2}$-$5P_{1/2}$-$5D$, of atomic rubidium are separated in the excitation space. 

In the rest of the paper, we first theoretically describe an imaging method using the USCC to revolve distinct nonlinear transition pathways in Sec.~II, and briefly summarize our experimental procedure in Sec.~III. The experimental results are presented in Sec.~IV, with a comparison of the performance of the three proposed modulations. 
The conclusion follows in Sec.~V.

\section{theoretical consideration}
As the simplest model, we consider a four-level system in the diamond-type configuration. There are two two-photon transition pathways,
 $|g\rangle \rightarrow |a\rangle \rightarrow |f\rangle$ and
$|g\rangle \rightarrow |b\rangle \rightarrow |f\rangle$, 
where ${|g\rangle}$ is the ground state, ${|a\rangle}$ and ${|b\rangle}$ are the intermediate states, and ${|f\rangle}$ is the final state. The energies are $0$, $\w_a$, $\w_b$, and $\w_f$ (in unit $\hbar$), respectively. From the second-order perturbation theory, the two-photon transition probability amplitude $c_{fg}$ to the final state $|f\rangle$ is given by
\begin{equation}
    c_{fg} = \sum_{i=a,b} \left( c_{fg,i}^{\rm r} + c_{fg,i}^{\rm nr} \right).
    \label{c_fg}
\end{equation}
Here, the resonant and non-resonant transition contributions are respectively defined as
\begin{align}
    c_{fg,i}^{\rm r} & = -\pi \frac{\mu_{fi} \mu_{ig}}{\hbar^2} 
    E(\w_{ig}) E(\w_{fi}), \\
    c_{fg,i}^{\rm nr} & = \I \frac{\mu_{fi} \mu_{ig}}{\hbar^2} 
    \int_{-\infty}^{\infty} \frac{ E \left(\w \right)
        E \left(\w_{fg} - \w \right)} {\w_{ig}-\w} d\w,
\end{align}
where $\mu_{ij}$ is the transition dipole moments and $\w_{ij}=\w_i-\w_j$ is the resonant transition frequencies~\cite{LeePRA2013}.
The $c_{fg,i}^{\rm r}$ term is the resonant transition contribution that only depends on the resonance frequency components of laser spectrum $E(\w)$. Due to the narrow spectral response in $c_{fg,i}^{\rm r}$, it is not possible to isolate the resonant excitation part in time (likewise in space). However, the $c_{fg,i}^{\rm nr}$ term, the non-resonant transition contribution, depends on all possible spectral pairs that satisfy the energy conservation $\w_i + \w_j = \w_{fg}$. Note that only $c_{fg,i}^{\rm nr}$ is involved in USCC~\cite{WJLeePRA2015}.

We first consider a single laser-pulse to be programmed in the frequency domain as
\begin{align}
    E_{\rm s}(t)
    &= \int E_{\rm s}(\w) e^{\I \w t} d\w 
    = \int A \left( \w \right) e^{\I  \Phi \left( \w \right) } e^{\I  \w t} d\w,
\end{align}
where $A(\w)$ and $\Phi(\w)$ are the programmed spectral amplitude and phase of the electric-field, respectively. In order to eliminate the resonant contribution $c_{fg,i}^{\rm r}$, we program spectral holes near the resonant frequencies $\w_{ag}$ and $\w_{bg}$, by
\begin{equation} 
    A (\w) = A_0(\w)  \left(1- e^{-{(\w-\w_{ag})^2}/{\delta_a^2}}
    - e^{-{(\w-\w_{bg})^2}/{\delta_b^2}}\right) .
    \label{mod_amp}
\end{equation}
Here, $A_0(\w)$ is the spectral amplitude before programming and $\delta_{i=a,b}$ is the width of each spectral hole, which is significantly smaller than the pulse bandwidth. With this spectral amplitude programming, the resonant transition background signal is removed (i.e., $c_{fg,i}^{\rm r}=0$),
so only the non-resonant transition part will be considered, as
\begin{equation}
     c_{fg} = c_{fg,a}^{\rm nr} + c_{fg,b}^{\rm nr} . 
    \label{eq:amplitude_removed}
\end{equation}
When two pulses of the same electric-field spectrum $E_s(\w)$ counter-propagate along $\pm z$, the combined electric field reads
\begin{equation}
E (\w)= A(\w)e^{\I \Phi(\w)} \left( e^{-\I \w z/c} +  e^{\I  \w z/c}\right),
\end{equation} 
and Eq.~\eqref{eq:amplitude_removed} can be replaced by
\begin{eqnarray} 
            c_{fg} \left( z \right)
            &=&  \sum_{i=a,b}
            \int_{-\infty}^{\infty}  \I  f_i(\w)
            A\left( \w \right) A\left( \hat{\w} \right)
            e^{ \I [\Phi \left(\w \right) + \Phi \left( \hat{\w} \right)] } \nonumber \\
            &\times&
            \left[ 1 +  e^{2\I \w_{fg}z/c} + e^{2\I \hat{\w} z/c} + e^{2\I  \w z/c} \right] d \w,
        \label{c_fg_detail}
    \end{eqnarray}
where for convenience we define $f_i(\w) = \mu_{fi} \mu_{ig} / [ \hbar^2 (\w_{ig}-\w)]$ and $\hat{\w}=\w_{fg}-\w$.
where for convenience we define $f_i(\w) = \mu_{fi} \mu_{ig} / [ \hbar^2 (\w_{ig}-\w)]$, $\hat{\w}=\w_{fg}-\w$ and the global phase factor $\exp(−i\w_{fg} z/c)$ is omitted.
Then, the spatial excitation probability is given by
\begin{widetext}
    \begin{align}
      |{c_{fg}}(z)|^2 
        = &\left \vert  \sum_{i=a,b}  \int_{-\infty}^{\infty}
        f_i(\w)
        A\left(\w \right) A\left( \hat{\w} \right)
        e^{ \I [\Phi \left(\w \right) + \Phi \left( \hat{\w} \right) ] }
        \left[ \cos \left( (\hat{\w} - \w)z/c \right) + 1 +  e^{2\I \w_{fg}z/c}  \right] d \w \right \vert ^2,
        \label{eq:S(z)}
    \end{align}
\end{widetext}
where the cosine term in the square bracket
corresponds to the counter-propagating pulse contribution, and the remaining two terms correspond to the single-sided pulse contribution.

The spectral phase programming in $ \Phi \left(\w \right) + \Phi \left( \hat{\w} \right)$ needs two strategies: a sign-flipping of the function $f_i(\w)$ at the intermediate resonances, and a maximizing of the ratio between the counter-propagating pulse and single-sided pulse contributions. After describing the experimental procedure in Sec.~III, we introduce three such phase-programming solutions in Sec.~IV, along with the corresponding experiments.

\section{experimental description}

Experiments were performed for the fine-structure transitions of atomic rubidium ($^{85}$Rb), where the four lowest energy levels are $|g\rangle=5S_{1/2}$, $|a\rangle=5P_{1/2}$, $|b\rangle=5P_{3/2}$, and $|f\rangle=5D$. A schematic of the experimental setup is shown in Fig.~\ref{fig1}, which is similar to the one from our earlier work~\cite{WJLeePRA2015}. We used a Ti:sapphire mode-locked laser oscillator, producing sub-picosecond laser pulses at a repetition rate of 82 MHz. 
The laser pulses were frequency-centered at $\w_0/2\pi = 384.3$~THz ($\w_0=\w_{fg}/2$ and equal to 778~nm in wavelength) to be two-photon resonant to the $5S_{1/2}$-$5D$ transition.
The laser bandwidth was $\Delta\w /2\pi = 22.3$ THz (with a full width at half maximum of 45~nm in wavelength), which was sufficient to cover all four transitions.

The laser pulse was programmed with a transmissive spatial light modulator (SLM) with an array of liquid-crystal pixels (128 pixels, 100~$\mu$m pitch) placed in the Fourier domain with a $4f$ geometry pulse shaper~\cite{SLM}. The focal length of the $4f$ geometry was $f=150$~mm, and the spectral resolution of each pixel was 0.46~nm in wavelength.
The groove density of the gratings was 1200~mm$^{-1}$. The spectral position of each SLM pixel was calibrated by scanning a $\pi$-step phase~\cite{LeePRA2013}. For the spectral amplitude modulation, two copper wires were placed on the Fourier plane: the first one was 140~$\mu$m in width to block the intermediate resonance $\omega_{ag}$, and the second one was 700~$\mu$m in width to block $\w_{bg}$ and also reduce the signal strength difference between the transition paths.
The as-programmed pulses were then focused in the rubidium vapor cell and the fluorescence at 420~nm from the $5D$ state through the $6P$ state was imaged with a charge-coupled device (CCD) camera.
The vapor cell was heated to around 50$\sim$60~$^{\circ}$C to enhance the fluorescence signal.

\section{Results}

\subsection{Double $V$-shape phase modulation}
\begin{figure*}[t]
\includegraphics[width=1.8\columnwidth]{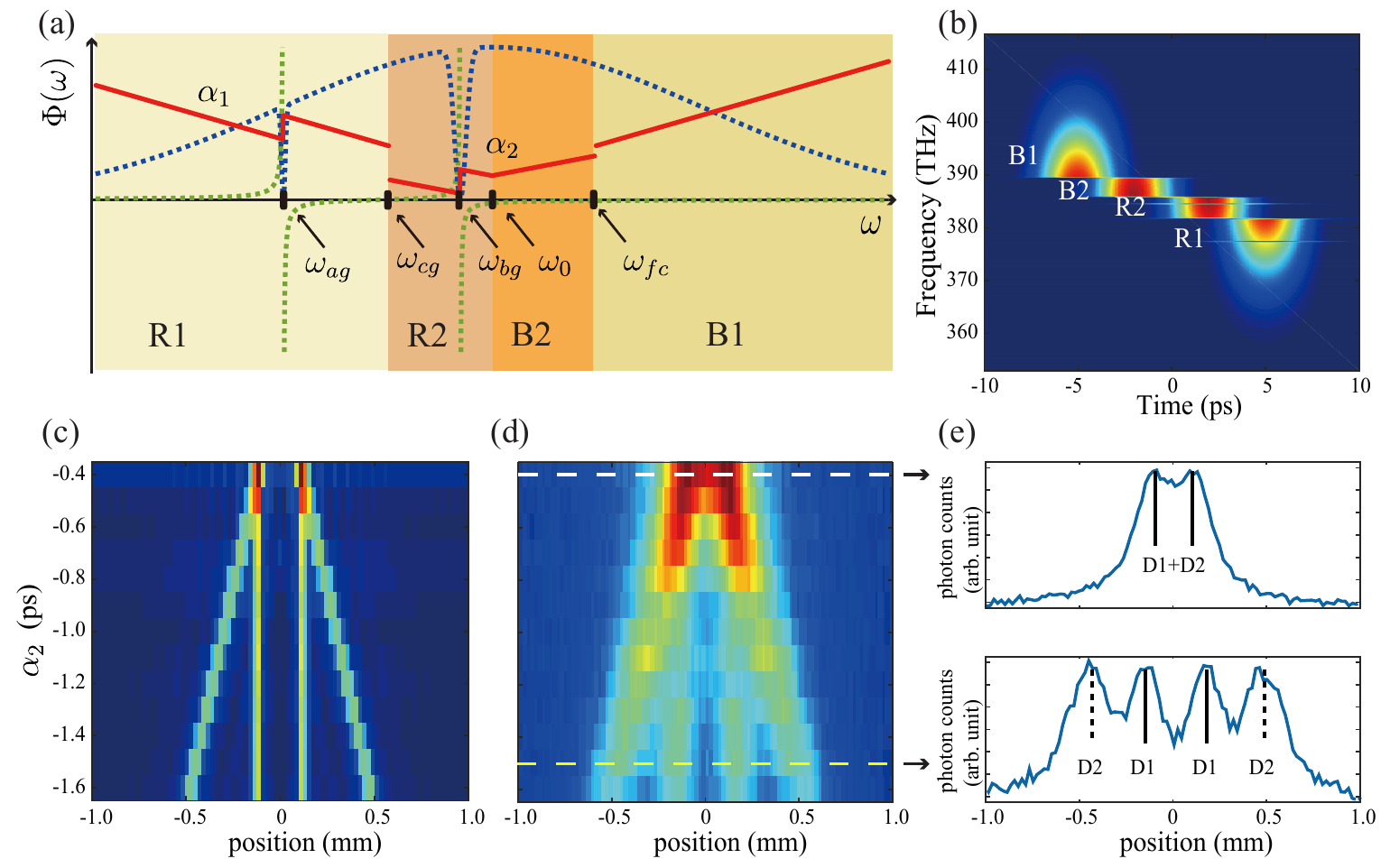}
\caption{(Color online)
    (a) The plot of double-$V$ shape spectral phase modulation (solid red line), modulated spectral amplitude (dotted blue line), and  $f_i(\omega)$ from Eq.~\eqref{eq:S(z)} (dotted green line).
    (b) Spectrogram of a pulse having the double-$V$ shape spectral phase from Eq.~\eqref{Double-V}.
    (c) Map of numerical calculation results of Eq.~\eqref{eq:S(z)} with the spectral phase modulation Eq.~\eqref{Double-V}, where $\alpha_1 = 0.4$ ps is fixed and $\alpha_2$ increases from $0.4$ ps to $1.6$ ps.
    (d) Composite map from experiment. (e) Experimental result for $\alpha_1 =  \alpha_2 = 0.4$ ps (upper) and for $\alpha_1 = 0.4$ ps, $\alpha_2 = 1.5$ ps (lower). The positions of $z_1$ (solid black lines) and $z_2$ (dashed black lines) are illustrated.
 \label{fig2}
 }
\end{figure*}

The first solution is a double $V$-phase modulation, a direct extension from Ref.~\cite{WJLeePRA2015}, which reads:
\begin{equation}
    \begin{aligned} 
        \Phi(\w)_{R1} &= -\alpha_1 (\w-\w_0) + \pi \Theta (\w-\w_{ag}) \\
        \Phi(\w)_{R2} &= -\alpha_2 (\w-\w_0) + \pi \Theta (\w-\w_{bg})  \\
        \Phi(\w)_{B2} &=  \alpha_2 (\w-\w_0) + \pi \\
        \Phi(\w)_{B1} &=  \alpha_1 (\w-\w_0) + \pi
    \end{aligned}
    \label{Double-V}
\end{equation}
where $\Theta(x)$ denotes the Heaviside step-function
and $\alpha_{1,2}$ are the spectral phase-slopes. The subscripts of $\Phi(\w)$ stand for the following spectral blocks: $R1=(0, \w_{cg})$, $R2= (\w_{cg}, \w_{0})$, $B1=(\w_{0}, \w_{fc})$, and $B2=(\w_{fc}, \infty)$. 
In Fig.~\ref{fig2}(a), the modulated spectral amplitude (dashed blue line) and phase (solid red line) are plotted, using Eqs.~\eqref{mod_amp} and \eqref{Double-V}, respectively, along with the inverse function $f_a(\w) + f_b(\w)$ (dotted green line).
In the given four-level system, with $\w_{bg} > \w_{ag}$, there are seven frequency boundaries to be considered: $\w_{ag}$, $\w_{cg}$, $\w_{bg}$, $\w_{0} (=\w_{fg}/2)$, $\w_{fb}$, $\w_{fc}$, and $\w_{fa}$, in ascending order. Here, $\w_{ag}$,  $\w_{bg}$, $\w_{fb}$, and $\w_{fa}$ are the resonances, and $\w_{cg}$ and $\w_{fc}$ are the characteristic frequencies where $f_a(\w) + f_b(\w)$ changes its sign~\cite{Cao2016,Cao2017}, given by
\begin{equation}
    \w_{cg} = \frac{k \w_{ag}+\w_{bg}} {k+1}, \quad
    k=\frac{\mu_{fb}\mu_{bg}}{\mu_{fa}\mu_{ag}},
\end{equation}
and $\w_{fc} = \w_{fg}-\w_{cg}$. The four spectral blocks are defined to each enclose the following characteristic resonances: $f_{ag} \in R1$, $f_{bg} \in R2$, $f_{fb}\in B2$, and $f_{fa} \in B1$. Then, due to the singular nature of $f_i(\w)$, the spectral region of $R1+B1$ dominantly contributes to the $5S_{1/2}$-$5P_{1/2}$-$5D$ (D1 transition) pathway and likewise, $R2+B2$ to $5S_{1/2}$-$5P_{3/2}$-$5D$ (D2). 

In the time domain, this phase modulation in Eq.~\eqref{Double-V} splits the initial unshaped pulse into four sub-pulses, each having a distinct spectral region. 
As shown in the spectrogram in Fig.~\ref{fig2}(b), these spectral blocks are time-shifted, with respect to the initial pulse, by $\Delta t = -\alpha_1$ ($B1$), $ -\alpha_2$ ($B2$), $\alpha_2$ ($R2$), and $\alpha_1$ ($R1$), when $\alpha_1>\alpha_2>0$. Then, the single-sided pulse contribution in Eq.~\eqref{eq:S(z)} is completely washed out, and the counter-propagating pulse contribution in Eq.~\eqref{eq:S(z)} is given with the integrand  $e^{2i\alpha(\omega-\omega_0)} \cos [ {\left( \hat{\w} - \w\right)z}/{c} ]$ that constructively interferes when $\alpha = {z}/{c}$. Since the shaped pulse, having four sub-pulses, meets its counter-propagating copy at the center of the vapor cell ($z=0$), four different two-photon transitions occur at positions $z_1 = \pm\alpha_1 c$ ($R1+B1$) and $z_2 = \pm\alpha_2 c$ ($R2+B2$). 

The numerical calculation and the experimental result for the double $V$-phase modulation are plotted in Fig.~\ref{fig2}(c) and (d), where $\alpha_2$ was increased from $0.4$ to $1.6$~ps with a step size of $0.1$~ps, while $\alpha_1=0.4$~ps was kept constant. For $\alpha_1 = \alpha_2 = 0.4$~ps, there are two sub-pulses split in the time domain, so two excitation peaks exist with overlapped D1 and D2 transitions, as shown in the upper part of Fig.~\ref{fig2}(e). As $\alpha_2$ increases, the sub-pulses having the spectral blocks corresponding to D2 move further from $z=0$, so the excitation peaks are better resolved as in the lower part of Fig.~\ref{fig2}(e).

\subsection{Three phase-slopes}

\begin{figure*}
\includegraphics[width=1.8\columnwidth]{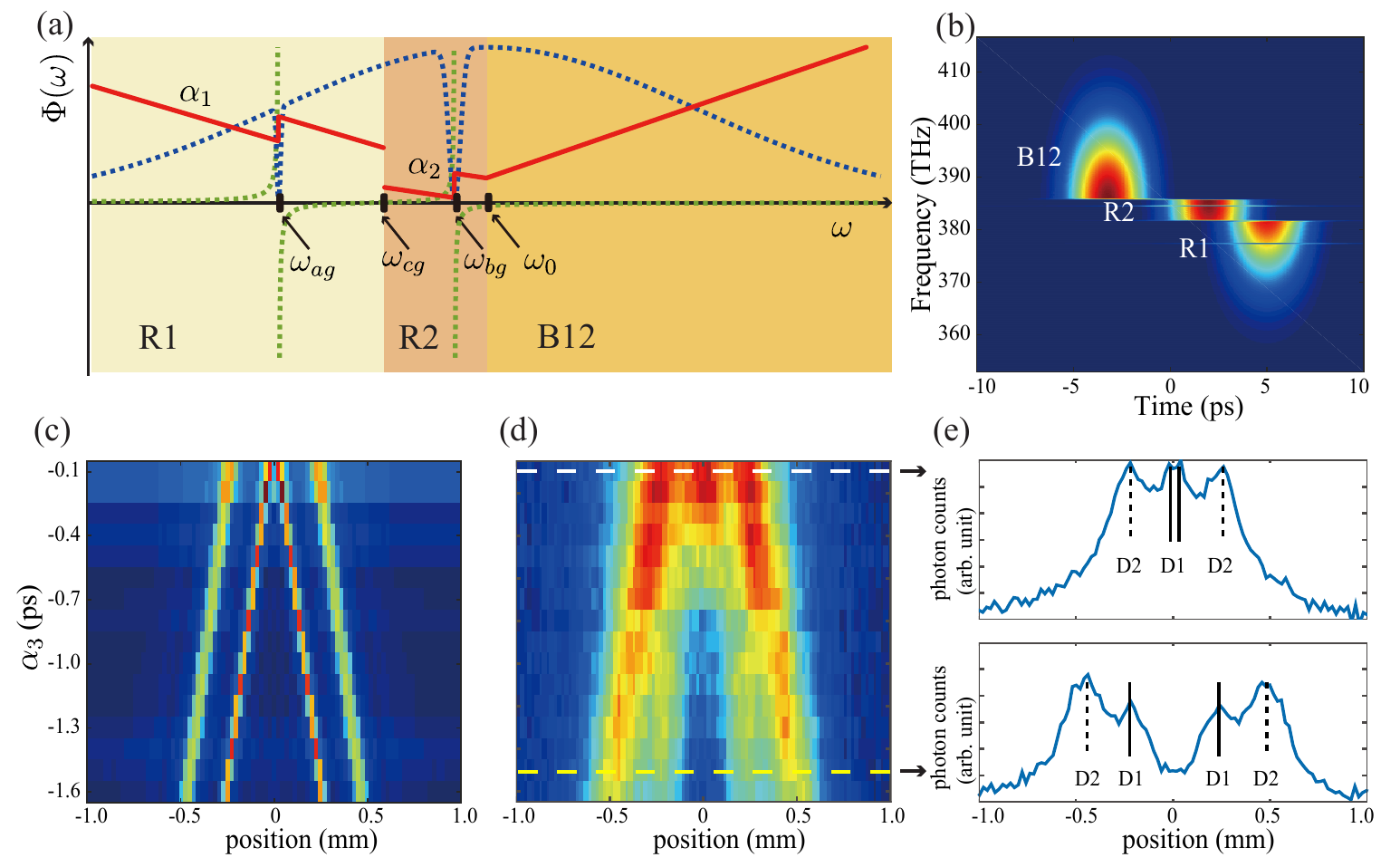}
\caption{(Color online) 
    (a) The plot of three phase slopes phase modulation (solid red line).
    (b) The spectrogram of a pulse having three phase slopes spectral phase from Eq.~\eqref{mod_3slopes}.
  (c) Map of numerical calculation results of Eq.~\eqref{eq:S(z)} with the spectral phase modulation Eq.~\eqref{mod_3slopes}, where $\alpha_1 = 0.1$ ps and $\alpha_2 = 1.5$ ps were fixed and $\alpha_3$ increased from $0.1$ ps to $1.6$ ps.
    (d) Composite map from experiment. (e) Experimental result for $\alpha_3 = 0.1$~ps and $\alpha_3=1.5$~ps, respectively. The positions of $z_1$(solid black lines) and $z_2$(dashed black lines) are illustrated.
}
 \label{fig3}
\end{figure*}

The second phase-programming solution utilizes three phase-slopes, which are given by:
\begin{equation}
    \begin{aligned}
        \Phi(\w)_{R1} &= -\alpha_1 (\w-\w_0) + \pi \Theta (\w-\w_{ag}) \\ 
        \Phi(\w)_{R2} &= -\alpha_2 (\w-\w_0) + \pi \Theta (\w-\w_{bg}) \\
        \Phi(\w)_{B12}&=  \alpha_3 (\w-\w_0) + \pi
    \end{aligned}
    \label{mod_3slopes}
\end{equation}
where $B12$ combines the spectral blocks $B1$ and $B2$. This phase modulation, plotted as the solid red line in Fig.~\ref{fig3}(a), results in three sub-pulses in the time-domain, as shown in the spectrogram in Fig.~\ref{fig3}(b). The first and third pulses (e.g., from the forward and backward propagating pulses, respectively) cause the two-photon transition through  D1, and the second and the third through D2.  As a result, the integrand of the counter-propagating pulse contribution term in Eq.~\eqref{eq:S(z)} becomes $e^{2i(\alpha_{1,2} + \alpha_3) (\omega-\omega_0)} \cos \left( {\left( \hat{\w} - \w\right)z}/{c} \right)$, and it constructively interferes when $\alpha_{1,2} + \alpha_3 = {2z}/{c}$. The excitation positions are determined as $z_1 = \pm (\alpha_1+\alpha_3)c/2$ and $z_2 = \pm (\alpha_2+\alpha_3)c/2$, respectively. The numerical calculation and experimental result are plotted in Figs.~\ref{fig3}(c) and (d), respectively, where $\alpha_1$ and $\alpha_2$ were fixed with $\alpha_1 = 0.1$~ps and $\alpha_2 = 1.5$~ps, while $\alpha_3$ was increased from 0.1 to 1.6~ps with a step size of 0.1~ps. In this case, the spacing between the excitation peaks for the respective D1 and D2 transitions is fixed because $\alpha_1$ and $\alpha_2$ are constants, while the peak positions are gradually separated as $\alpha_3$ increases. 

\subsection{Periodic square phase}
\begin{figure}
\includegraphics[width=1.0\columnwidth]{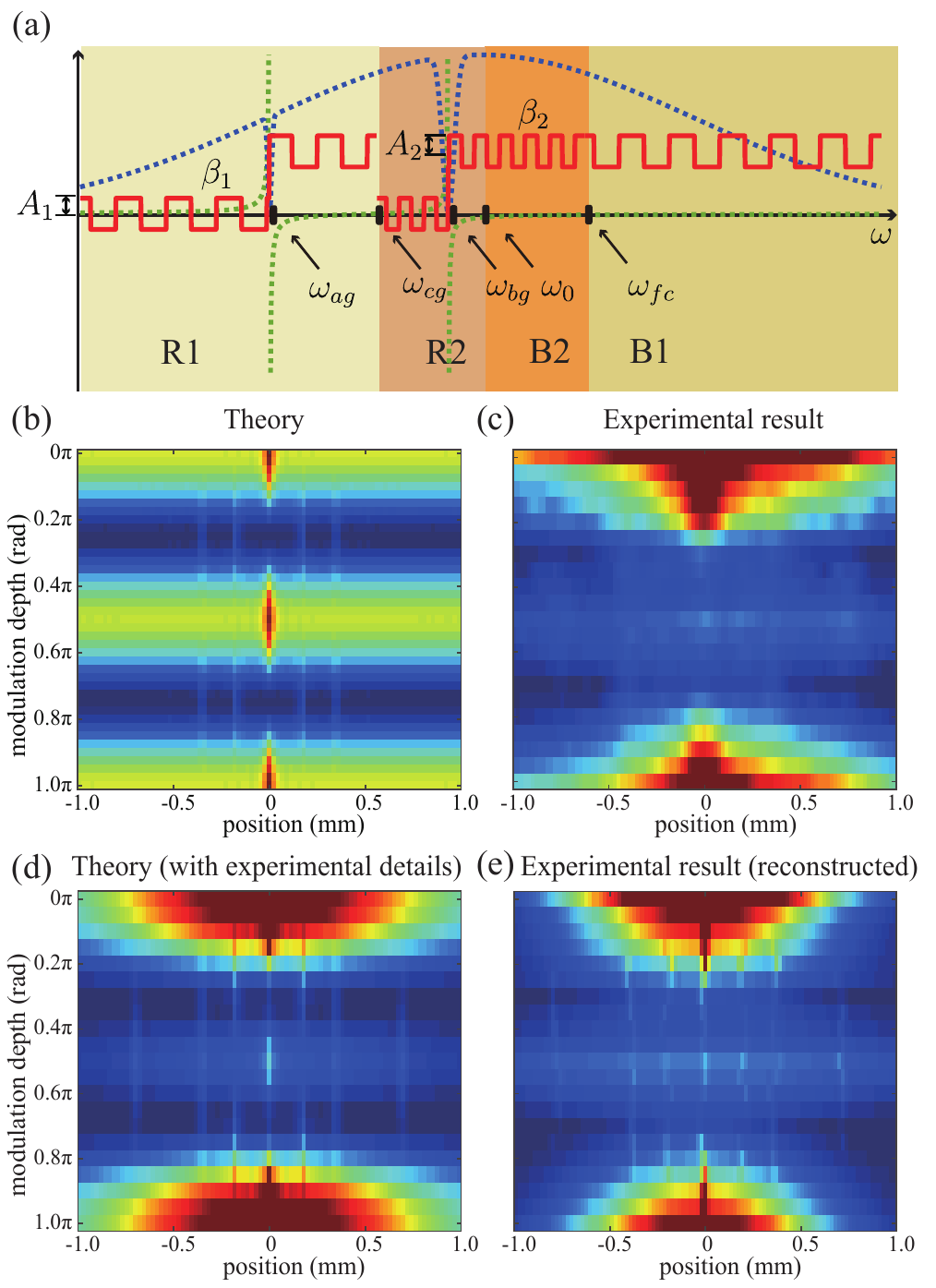}
\caption{(Color online) 
    (a) Plot of periodic square spectral phase modulation.
   (b) Numerical calculation result of Eq.~\eqref{eq:S(z)} with the spectral phase modulation Eq.~\eqref{square_phase}, changing the modulation depth from $A_1 = A_2 = 0 $ to $\pi$.
    (c) Composite map from experiment.
    (d) Numerical calculation including finite SLM pixel size and intensity distribution by beam focusing.
    (e) Reconstructed result by eliminating the atomic motion effect.
 }
\label{fig4}
\end{figure}

The last phase modulation solution utilizes the mathematical relation
$e^{i\frac{\pi}{2}[\mathrm{sgn}(x)-1]}\cos(x)= |\cos(x)|$ to include the counter-propagating pulse contribution in Eq.~\eqref{eq:S(z)} (the first term in the square bracket), with all spectral pairs in-phase. The as-obtained solution reads:
The last phase modulation solution utilizes the mathematical relation $e^{i\frac{\pi}{2}[\mathrm{sgn}(\cos x)-1]}\cos x= |\cos x|$ to add the counter-propagating pulse contribution in Eq.~\eqref{eq:S(z)} (the first term in the square bracket) all in phase.
The as-obtained solution reads:
\begin{equation}
    \begin{aligned}
        \Phi_{R1}(\w) &= {A_1} \mathrm{sgn} \left[ \cos({\beta_1} (\w -\w_0)) \right] + \pi \Theta (\w-\w_{ag})  \\
        \Phi_{R2}(\w) &= {A_2} \mathrm{sgn} \left[ \cos ( \beta_2 (\w -\w_0) ) \right] + \pi \Theta (\w-\w_{bg}) \\
        \Phi_{B2}(\w) &= {A_2} \mathrm{sgn} \left[ \cos({\beta_2} (\w -\w_0)) \right] + \pi  \\
        \Phi_{B1}(\w) &= {A_1} \mathrm{sgn} \left[ \cos ( \beta_1 (\w -\w_0) ) \right] + \pi 
        \label{square_phase}
    \end{aligned} 
\end{equation}
where $A_{1,2}$ are the modulation amplitudes, $\mathrm{sgn}(x)$ is the signum function defined as $\mathrm{sgn}(x) = -1$ for $x\leq 0$ and $+1$ for  $x>0$, and $\beta_{1,2}$ are the modulation frequencies, or the period, of the square function for each spectral region. 
When the modulation amplitudes and frequencies satisfy $A_1 = A_2 = \pi/4$ and $\beta_{1,2} = {2z}/{c}$, as illustrated in Fig.~\ref{fig4}(a), the integrand in Eq.~\eqref{eq:S(z)} becomes $e^{i\frac{\pi}{2}\mathrm{sgn}[\cos(2(\omega-\omega_0)z/c)]} \cos [2({\w}_0 - \w)z/c ]=e^{i\pi/2}|\cos(2(\omega_0-\omega)z/c)|$, which induces a complete constructive interference from the counter-propagating pulse contributions.
As a result, excitation occurs at $z_1 = \pm{\beta_1 c}/{2}$ and $z_2 = \pm{\beta_2 c}/{2}$ (the spatially resolved two-photon excitation pattern). 
On the other hand, when $A_1 = A_2 = \pi/2$, the integral of the counter-propagating pulse contribution term is eliminated because $e^{i\pi\mathrm{sgn}[\cos(\beta_{1,2}(\omega-\omega_0))]} \equiv -1$, and only the single-sided excitation term (the spatially independent signal) remains in Eq.~\eqref{eq:S(z)}. 
Therefore, there is no specific spatial excitation pattern. The numerical calculation of Eq.~\eqref{eq:S(z)} with the solution of Eq.~\eqref{square_phase} is shown in Fig.~\ref{fig4}(b), where the modulation depth is changed from $A_1 = A_2 = 0$ to $A_1 = A_2 = \pi$ with $0.025\pi$ steps for $\delta=0$, with $\alpha_1 = 1.16$ ps and $\alpha_2 = 2.32$ ps.

The experimental result for the periodic square phase solution is shown in Fig.~\ref{fig4}(c), where $A_{1,2}$ were scanned together from $A_1 = A_2 = 0$ to $\pi$ with a step of $0.05\pi$. The optimal modulation amplitude with which the single-side contribution is minimized was found to be $A_1 = A_2 = 0.35\pi$, slightly shifted from the theoretically predicted $A_1 = A_2 = \pi/4$. This discrepancy mainly comes from the finite size of the phase modulation pixels of our SLM, as the  modulation functions of two transition pathways ($\Phi_1$ and $\Phi_2$ in Eq.~\eqref{square_phase}) could not both be centered on the two-photon center $\omega_0$ simultaneously. Another discrepancy originates from the spatial background intensity distribution by beam focusing. This is more apparent for the modulation depths where the single-side contribution is dominant. The combined numerical calculation with these constructive interference leakage effects in the experimental system is illustrated in Fig.~\ref{fig4}(d), which better matches the result in Fig.~\ref{fig4}(c). In spite of these systematic errors, the overall behavior where the single-sided contribution is recovered at $A_1 = A_2 = \pi/2$ and the minimized point exists between $A_1 = A_2 = 0$ and $\pi/2$ remains.

The effects of atomic motion in vapor and the imaging blurring due to Abbe diffraction are taken into account to reconstruct Fig. \ref{fig4}(e). To investigate the extent of each excitation peak broadening, the convolution integral of the numerical calculation results of excitation patterns with these effects were performed. For the atomic motion effects, the position distribution along the pulse propagation axis after a $5D$ state lifetime of 240~ns from the velocity distribution of Rubidium (for the pulse propagation axis) at a temperature 55$^\circ$C was considered. For the diffraction effect from the imaging aperture, an impulse response function with a lens aperture size of 25.4~mm was taken into account. The width of the broadened excitation peak after the convolution integral with atomic motion effects was approximately 120~$\mu$m, which is comparable to the average width of each excitation pattern in the experimental result in Fig.~\ref{fig4}(c) with Gaussian fitting ($R$-square above 0.965). The width of the broadened excitation peak by diffraction was on the order of a few~$\mu$m, which is smaller than the pixel width of our CCD camera, and therefore negligible. Then, comparing the width of the Gaussian fittings of excitation patterns ($R$-square above 0.99) before and after convolution, we found that this broadened excitation pattern could be effectively reconstructed by multiplying a Gaussian with a FWHM of approximately 30~$\mu$m to each broadened excitation peak.  By Gaussian fitting and multiplying the Gaussian function found above, the broadening effect of the experimental results in Fig. \ref{fig4}(c) is effectively reconstructed. This result is shown in Fig.~\ref{fig4}(e). Each excitation peak narrowed, agreeing well with the measurement in Fig.~\ref{fig4}(d).

\section{Conclusion}
In summary, we performed ultrafast spatial coherent-control experiments to resolve the fine-structure two-photon transitions  ($5S_{1/2}$-$5P_{1/2}$-$5D$ and $5S_{1/2}$-$5P_{3/2}$-$5D$ pathways) of atomic rubidium with various phase programming solutions. This work not only extended our earlier work~\cite{WJLeePRA2015} of combining spectral amplitude and phase programmings to deal with atomic transitions with multiple intermediate resonances, but also newly proposed and demonstrated two additional phase programming solutions. Compared to the previously introduced double-$V$ spectral phase, the three phase slopes and the periodic square phases provided for simpler programming and the possibility for finer spectral resolution spectroscopy, respectively.
In experiment, counter-propagating ultrafast optical pulses as-shaped with the spectral phase and amplitude programming solutions successfully induced the given two-photon transitions, simultaneously and at distinct spatial locations, agreeing well with the theoretical analysis.

\begin{acknowledgments}
This research was supported by Samsung Science and Technology Foundation [SSTF-BA1301-12]. The authors thank Geol Moon and Hangyeol Lee for useful discussion. D. Cao, F. Gao and F. Shuang acknowledge support from the National Natural Science Foundation of China (Grants No. 61720106009, No. 61773359, No. 61403362 and No.61374091). F. Shuang thanks the Leader talent plan of the Universities in Anhui Province and the CAS Interdisciplinary Innovation Team of the Chinese Academy of Sciences for financial support.
\end{acknowledgments}


\begin{thebibliography}{99}


\bibitem{Bergmann} K. Bergmann, H. Theuer, and B. W. Shore,
``Coherent population transfer among quantum states of atoms and molecules," 
Rev. Mod. Phys.~{\bf 70}, 1003-1025 (1998).

\bibitem{Shapiro} M. Shapiro and P. Brumer, {\it Principles of the
Quantum Control of Molecular Processes} (Wiley, New York, 2003).

\bibitem{Tanner} D.J. Tanner and S. A. Rice, 
``Control of selectivity of chemical reaction via control of wavepacket evolution," 
J. Chem. Phys.~{\bf 83}, 5013-5018 (1985).

\bibitem{Dudovich} N. Dudovich, B. Dayan, S. M. Gallagher Faeder, and
Y. Silberberg, 
``Transform-limited pulses are not optimal for resonant multiphoton transitions," 
Phys. Rev. Lett. {\bf 86}, 47-50 (2001).

\bibitem{SLee} S. Lee, J. Lim, and J. Ahn,
"Strong-field two-photon absorption in atomic Cesium: an analytic approach,"
Opt. Express {\bf 17}, 7648 (2009).

\bibitem{LeePRA2013} H. G. Lee, H. Kim, J. Lim, and J. Ahn,
``Quantum interference control of four-level diamond-configuration quantum system,''
Phys. Rev. A {\bf 88}, 053427 (2013).

\bibitem{MCStowe} M. C. Stowe, A. Pe'er, and J. Ye,
``Control of Four-Level Quantum Coherence via Discrete Spectral Shaping of an Optical Frequency Comb,"
Phys. Rev. Lett.~{\bf 100}, 203001 (2008).

\bibitem{Dark_pulses} D. Meshulach and Y. Silberberg, 
``Coherent quantum control of two-photon transitions by a femtosecond laser pulse,'' 
Nature {\bf 396}, 239 (1998).

\bibitem{Teaching_laser} R. S. Judson, H. Rabitz,
"Teaching lasers to control molecules,"
Phys. Rev. Lett. {\bf 68},
1500-1503 (1992).

\bibitem{Quantum_gates} J. J. Garc\'{i}a-Ripoll, P. Zoller, and J. I. Cirac, ``Speed Optimized Two-Qubit Gates with Laser Coherent Control Techniques for Ion Trap Quantum Computing,'' 
Phys. Rev. Lett. {\bf 91}, 157901 (2003).

\bibitem{HHG} X. Wang, C. Jin, and C. D. Lin, ``Coherent control of high-harmonic generation using waveform-synthesized chirped laser fields,'' 
Phys. Rev. A {\bf 90}, 023416 (2014).

\bibitem{NDudovich} N. Dudovich, D. Oron, and Y. Silberberg,
``Single-pulse coherently controlled nonlinear Raman spectroscopy and microscopy,"
Nature~{\bf 418}, 512 (2002).

\bibitem{BarmesNPhoton2013} I. Barmes, S. Witte, and K. S. E. Eikema, 
``Spatial and spectral coherent control with frequency combs,'' 
Nat. Photonics {\bf 7}, 38 (2013).
% First demonstration of Spatial and spectrla coherent control with v-type, comb

\bibitem{JonesScience2000} D. J. Jones, S. A. Diddams, J. K. Ranka, A. Stentz, R. S. Windeler, J. L. Hall, and S. T. Cundiff, 
``Carrier-envelope phase control of femtosecond mode-locked lasers and direct optical frequency synthesis,'' 
Science  {\bf 288}, 635 (2000). 
% frequency comb science paper

\bibitem{HolzwarthPRL2000} R. Holzwarth, Th. Udem, T. W. H{\"a}nsch, J. C. Knight, W. J. Wadsworth, and P. St. J. Russell, 
``Optical frequency synthesizer for precision spectroscopy,'' 
Phys. Rev. Lett. {\bf 85}, 2264 (2000).
% frequency comb PRL paper

\bibitem{BarmesPRL2013} I. Barmes, S. Witte, and K. S. E. Eikema, 
``High-precision spectroscopy with counterpropagating femtosecond pulses,'' 
Phys. Rev. Lett. {\bf 111}, 023007 (2013).

\bibitem{WJLeePRA2015} W. Lee, H. Kim, K. Kim, and J. Ahn, ``Coherent control of resonant two-photon transitions by counter-propagating ultrashort pulse pairs,'' 
Phys. Rev. A {\bf 92}, 033415 (2015).

\bibitem{Gao2014} F. Gao, R. Rey-de-Castro, A.M. Donovan, J. Xu, Y. Wang, H. Rabitz, F. Shuang, 
    ``Pathway dynamics in the optimal quantum control of rubidium: Cooperation and competition,''
    Phys. Rev. A {\bf 89}, 023416 (2014).

\bibitem{Gao2015} F. Gao, Y. Wang, R. Rey-de-Castro, H. Rabitz, F. Shuang,
    ``Quantum control and pathway manipulation in rubidium,''
    Phys. Rev. A {\bf 92}, 033423 (2015). 

\bibitem{Gao2016} F. Gao, R. Rey-de-Castro, Y. Wang, H. Rabitz, F. Shuang, 
    ``Identifying a cooperative control mechanism between an applied field and the environment of open quantum systems,''
    Phys. Rev. A {\bf 93}, 053407 (2016).

\bibitem{SLM} A. M. Weiner, ``Femtosecond pulse shaping using spatial light modulators,'' Rev. Sci. Instrum. {\bf 71}, 1929 (2000).

\bibitem{Cao2016} D. Cao, L. Yang, Y. Wang, F. Shuang, F. Gao, 
    ``Controlling pathway dynamics of a four-level quantum system with pulse shaping,''
    J. Phys. A Math. Theor. {\bf 49}, 285302 (2016).

\bibitem{Cao2017} D. Cao, Y. Wang, S. Li, L. Yang, F. Shuang, and F. Gao,
    ``Optimal control of multiple two-photon transitions,''
    J. Math. Chem. {\bf 55}, 1053-1066 (2017).

\end{thebibliography}
\end{document}